\documentstyle[12pt,psfig]{article}
\newcommand{\bold}[1]{\mbox{\boldmath $#1$}} 

\newcommand{\mywarn}{******MY WARNING****** : }

\newcommand{\cref}{$\clubsuit$ \marginpar{$\clubsuit$ {\em Check ref.!}}
\typeout{\mywarn Check reference(s)!}} 

\def\ie{{\em i.e., }}

\def\be{\begin{equation}}
\def\ee{\end{equation}}
\def\br{\begin{eqnarray}}
\def\er{\end{eqnarray}}
\def\brn{\begin{eqnarray*}}
\def\ern{\end{eqnarray*}}
\def\rf#1{{(\ref{#1})}}

\def\Ket#1{||#1 \rangle}
\def\Bra#1{\langle #1||}
\begin{document}
\title{Rho-Nucleon Tensor Coupling and
Charge-Exchange Resonances}
\author{C. De Conti  \\[-.8ex]
{\normalsize \it Departamento de F\'{\i}sica, 
Instituto Tecnol\'{o}gico de Aeron\'{a}utica,} \\[-.8ex]
{\normalsize \it Centro T\'{e}cnico Aeroespacial,} \\[-.8ex]
{\normalsize \it 12228-900 S\~{a}o Jos\'{e} dos Campos, SP, Brazil} \\
 A. P. Gale\~{a}o  \\[-.8ex]
{\normalsize \it Instituto de F\'{\i}sica Te\'{o}rica,
 Universidade Estadual Paulista,} \\[-.8ex]
{\normalsize \it Rua Pamplona 145,
 01405-900 S\~{a}o Paulo, SP, Brazil} \\  
and \\ F. Krmpoti\'{c} \\[-.8ex]
{\normalsize \it Departamento de F\'{\i}sica, 
Facultad de Ciencias Exactas,} \\[-.8ex]
{\normalsize \it Universidad Nacional de La Plata,} \\[-.8ex] 
{\normalsize \it C.C. 67, 1900 La Plata, Argentina}}
\date{13 March 2000\\
{\normalsize Revised 15 July 2000} }
\maketitle

\vspace{-1\baselineskip}
\begin{abstract}
The Gamow-Teller resonance in $^{208}$Pb  is  discussed in the context
of a self-consistent RPA, based on the relativistic mean field theory.
We  inquire on the possibility of
substituting the phenomenological Landau-Migdal force by a microscopic
nucleon-nucleon interaction, generated from the rho-nucleon tensor coupling.
The effect of this coupling turns out to be very small when
the short range
correlations are not taken into account, but too large when these
correlations are simulated by the simple extraction of the contact
terms from  the resulting nucleon-nucleon interaction.

\vspace{1\baselineskip} 
\small \noindent 
{\it PACS:} 21.60.-n; 21.60.Jz; 21.30.Fe; 24.30.Cz \\
{\it Keywords:} Relativistic mean field theory; Random phase approximation; 
One-boson-exchange models; Charge-exchange resonances
\end{abstract}
\newpage
The quantum hadrodynamics (QHD) aims to describe the nuclear many-body
system in terms of nucleons and mesons \cite{Ser86}.
Proposed initially as a full-fledged renormalizable quantum field theory,
nowadays it is seen as an {\em effective field theory}, derivable,
in principle, from the quantum chromodynamics \cite{Ser97}.

The relativistic mean field theory (RMFT), which can be thought as a mean
field (Hartree)  approximation to the QHD, has been applied with great
success during the last few decades.
For instance, it accounts for both i) the nuclear matter saturation, and
ii) the ground state properties of finite nuclei along the whole periodic
table  \cite{Gam90}.
More recently, the RMFT has also been exploited for the description of unstable
nuclei all up to the nucleon drip lines \cite{Rin96}.

Through a relativistic version of the random phase approximation (RRPA),
various excited states and resonances have been studied in the context of the
RMFT \cite{Blu88}--\cite{Ma97} as well.
Quite recently we have also reported \cite{Con98} the  first calculation of
this type for the Gamow-Teller (GT) and isobaric analogue (IA) resonances,
excited from the ground states of $^{48}$Ca, $^{90}$Zr and
$^{208}$Pb nuclei.

Because of its pseudoscalar nature, the pion does not participate in the
description of the the ground states in the RMFT. Thus, besides the
nucleon and the Coulomb fields, only the $\sigma$, $\omega$ and $\rho$
mesons are usually involved in the calculations.
Yet, in dealing with isovector excitations it is essential to
include, together with the $\rho$ meson, the $\pi$ meson as well.
This has already been done in our previous work \cite{Con98},
with the pseudovector pion-nucleon coupling $f_\pi$ fixed at its
experimental value. For the remaining mesons, only the nonderivative
couplings to the nucleon were included, as usually done in RMFT.
With this prescription we were not able to
reproduce the excitation energies of the just mentioned
resonances. This has been possible only after introducing the
repulsive Landau-Migdal (LM) delta force
\begin{eqnarray}
 V_{LM}(1,2) &=& g' \left( \frac{f_{\pi}}{m_{\pi}} \right)^2
\bold{\tau}_1 \cdot \bold{\tau}_2 \; \bold{\sigma}_1 \cdot
\bold{\sigma}_2 \; \delta (\bold{r}_1 - \bold{r}_2), \label{1}
\end{eqnarray}
of the same magnitude ($g'=0.7$)  as the one used in the nonrelativistic
calculations \cite{Kre81}.

Here we wish to analyze whether the tensor (derivative) coupling
of the $\rho$ meson to the nucleon could generate a sufficiently
repulsive nucleon-nucleon force in order to locate the GT
resonance at the correct experimental energy and in this way
substitute the phenomenological LM force.
The IA resonance is practically not affected by this part of the
$\rho$-meson-exchange
potential and therefore it will not be discussed so exhaustively
as we do with the GT resonance.

 As mentioned above, it is not usual to include the tensor coupling of
the vector mesons to the nucleon in RMFT. This is because its
effect on the ground state is (rightly) thought to be small. On a
more general perspective, however, there are two good reasons why
one should do so. For one, according to the rules of effective
field theory such terms should appear in the effective QHD
Lagrangian \cite{Fur95}. For another, and perhaps more important
reason for the phenomenological stand we are taking, it is well
known that the tensor $\rho$-nucleon coupling gives a large
contribution to the spin-isospin component of the nucleon-nucleon
interaction \cite{Mac89}, and as such it could have an important
effect on the dynamics of the GT  resonance.

Our Lagrangian density is now
\begin{eqnarray}
{\cal L} & = & \bar{\psi} (i\gamma_{\mu} \partial^{\mu} - M)\psi \nonumber \\    
     &   & \mbox{} + \frac{1}{2} \partial_{\mu} \sigma \partial^{\mu} \sigma
               - \frac{1}{2} {m_{\sigma}}^{2} \sigma^{2} - \frac{1}{3} g_{2}
              \sigma^{3} - \frac{1}{4} g_{3} \sigma^{4} - g_{\sigma} \bar{\psi}
               \psi \sigma \nonumber \\
         &   & \mbox{} - \frac{1}{4} \Omega_{\mu\nu} \Omega^{\mu\nu}
               + \frac{1}{2} {m_{\omega}}^{2} \omega_{\mu} \omega^{\mu}
               - g_{\omega} \bar{\psi} \gamma_{\mu} \psi \omega^{\mu} \nonumber
\\
         &   & \mbox{} + \frac{1}{2} \partial_{\mu} \bold{\pi} \cdot
               \partial^{\mu} \bold{\pi} - \frac{1}{2} {m_{\pi}}^{2} \bold{\pi}
               \cdot \bold{\pi} - \frac{f_{\pi}}{m_{\pi}} \bar{\psi} \gamma_{5}
               \gamma_{\mu} \bold{\tau} \psi \cdot \partial^{\mu} \bold{\pi}
\nonumber \\
         &  & \mbox{} - \frac{1}{4} {\bold{R}}_{\mu\nu} \cdot \bold{R}^{\mu\nu}
               + \frac{1}{2} {m_{\rho}}^{2} \bold{\rho}_{\mu} \cdot
               \bold{\rho}^{\mu} - g_{\rho} \bar{\psi} \gamma_{\mu}
               \bold{\tau} \psi \cdot \bold{\rho}^{\mu} \nonumber \\
         &   & \mbox{} - \frac{f_{\rho}}{2M}\bar{\psi}\sigma_{\mu\nu}
               \bold{\tau}\psi\cdot\partial^{\mu}\bold{\rho}^{\nu} \nonumber \\
         &   & \mbox{} - \frac{1}{4} F_{\mu\nu} F^{\mu\nu} - e \bar{\psi}
               \gamma_{\mu} \frac{1+\tau_{3}}{2} \psi A^{\mu},
\label{2}
\end{eqnarray}
where
\begin{eqnarray}
\Omega^{\mu\nu} & = & \partial^{\mu}
\omega^{\nu}-\partial^{\nu}\omega^{\mu}, \nonumber\\
\bold{R}^{\mu\nu} & = & \partial^{\mu} \bold{\rho}^{\nu} -
\partial^{\nu} \bold{\rho}^{\mu} - 2g_{\rho} \bold{\rho}^{\mu}
\times \bold{\rho}^{\nu}, \label{3}
\\ F^{\mu\nu} & = &
\partial^{\mu} A^{\nu} - \partial^{\nu} A^{\mu}. \nonumber
\end{eqnarray}
This Lagrangian is identical to that of reference \cite{Con98}, except
for the $\rho$-nucleon tensor coupling term (the one proportional to
$f_{\rho}$).\footnote{There is a minor correction to be made in \cite{Con98}. 
One must replace $g_{\rho}$ by $2g_{\rho}$ in eqs. (1) and (2) of that reference 
for consistency with the remaining equations.}
Therefore, following the same route one arrives at identical
equations for the mean boson fields, except for that of the $\rho$ meson (only
the component $\rho^{0}_{3}$ survives for spherical, definite-charge nuclei),
which now takes the form
\begin{equation}
\left( - \nabla^2 + {m_\rho}^2 \right) \rho_3^0 = g_\rho {\rho_3}(r)
  + \frac{f_\rho}{2M} \nabla \cdot {\bold{\rho}_{t3}}(r),
\label{4}
\end{equation}
where the (vector) isovector density $\rho_3$ is as defined in
\cite{Con98} and we have introduced the tensor isovector density
\begin{equation}
\bold{\rho}_{t3} =
 \left \langle \bar{\psi} i \bold{\alpha} \tau_3 \psi \right \rangle =
 \sum_{\alpha=1}^{A} \bar{\cal U}_\alpha i\bold{\alpha} \tau_3 {\cal U}_\alpha.
\label{5}\end{equation}
The summation is over all the occupied single-particle, positive-energy states
${\cal U}_\alpha$, which obey the mean-field Dirac equation.
This is also modified to
\begin{eqnarray}
\left\{ - i \bold{\alpha} \cdot \nabla  + \beta \left[ M + {V_s}(r) \right]
+ {V_v}(r)
\mbox{} + ( i \beta \bold{\alpha} \cdot \bold{r}/r )
 {V_t}(r) \right\} {\cal U}_\alpha & = & E_\alpha {\cal U}_\alpha .
\nonumber\\
\label{6}
\end{eqnarray}
Again the scalar ($V_s$) and vector ($V_v$) potentials are as defined in
\cite{Con98}, while the tensor potential,
\begin{equation}
V_t = - \frac{f_\rho}{2M} \frac{d\rho^0_3}{dr} \tau_3,
\label{7}\end{equation}
is the contribution from the tensor-coupling term in (\ref{2}).

The general structure and derivation of the RRPA for
charge-exchange excitations, in the discretized spectral version
we use, has been delineated in \cite{Con98}. An alternative, more
detailed account can be found in \cite{Con99}. The main ingredient
is the residual interaction $V$. For a self-consistent
calculation, this must be obtained from the same Lagrangian
(\ref{1}) used for the mean field. Also, since Fock terms are
ignored in RMFT, we must consider only the {\em direct} matrix
elements of $V$. Hence only the isovector mesons contribute, and
we get $V = V_\pi + V_\rho\,$, with, in the instantaneous
approximation,

\begin{eqnarray}
V_\pi(1,2) &=& - \left(\frac{f_{\pi}}{m_{\pi}}\right)^2
\bold{\tau}_1\cdot\bold{\tau}_2 \; (\bold{\sigma}_1\cdot\nabla_1
\; \bold{\sigma}_2\cdot\nabla_2) Y(m_\pi,r_{12}),
\label{8}\end{eqnarray}
\newpage
\begin{eqnarray}
V_\rho(1,2) &=& \bold{\tau}_1 \cdot \bold{\tau}_2 \; \left[ \left(
g_\rho - \frac{f_\rho}{2M} i \beta \bold{\alpha} \cdot \nabla
\right)_1 \left( g_\rho - \frac{f_\rho}{2M} i \beta \bold{\alpha}
\cdot \nabla \right)_2 \right. \nonumber \\ & & \left. \mbox{} -
\left( g_\rho \bold{\alpha} + \frac{f_\rho}{2M} \beta
\bold{\sigma} \times \nabla \right)_1 \cdot \left( g_\rho
\bold{\alpha} + \frac{f_\rho}{2M} \beta \bold{\sigma} \times
\nabla \right)_2 \right]
Y(m_\rho,r_{12}),\nonumber \\ \label{9}
\end{eqnarray}
where $r_{12} = \left| \bold{r}_1 - \bold{r}_2 \right|$ and
$Y(m,r)= \exp(-mr)/(4\pi r)$.

For the numerical values of the parameters we follow mostly the
philosophy of \cite{Con98}, adopting the parameter set NL1
\cite{Gam90,Rei86}. Yet,  in view of the  difficulties
encountered by Ma {\it et al.} \cite{Ma97} in accounting for
the E1 and E0 giant resonances with the NL1 parameters,
a few results for the  TM1 model, worked out by Sugahara and
Toki \cite{Sug94}, will be presented as well.\footnote{In the latter case, the 
$\omega$-meson self-interaction term, not appearing in \rf{2}, was also included 
in the numerical calculations.} 
Taking experimental values for the pion,
the only new parameter is the $\rho$-nucleon tensor coupling
constant $f_\rho$. As mentioned in \cite{Bou87}, the vector
dominance model predicts for the ratio $f_\rho/g_\rho\equiv K_\rho$ a
value equal to the isovector magnetic moment of the nucleon, {\it
i.e.}, $\mu_p - \mu_n - 1 = 3.7$. On the other hand, most
meson-exchange models for the nuclear force use $K_\rho=6.6$
\cite{Mac89}. The former choice was preferred in the description
of the ground state properties in closed shell nuclei within  the
relativistic Hartree-Fock approximation \cite{Bou87}. Thus, the
discussion that follows will mainly rely on the lower value for
$K_\rho$, even though we are aware of the fact that the inclusion of
Fock terms can considerably change the adjusted values of the QHD
parameters \cite{Bou87}.

Another point to consider is whether the inclusion of the tensor
coupling term in the Lagrangian (\ref{2}) does not sensitively
affect the values of the remaining parameters. Fortunately, while
the contribution of this term is not strictly zero in RMFT, its
effects on the single particle energies as well as on the ground
state properties are certainly very small. We therefore feel justified
in keeping the remaining parameters fixed at their NL1 or TM1 values.
With $K_\rho=3.7$,  for instance, the spin-orbit splitting is modified in
less than $150$ KeV.
\footnote{For identical particles, the NL1 paramerization yields significantly 
larger spin-orbit splittings than the TM1 model, while the opposite happens for 
nonidentical particles.}
Similarly tiny effects on the energy per particle
and the root-mean-square radii are displayed in Table \ref{t1}.
An interesting side remark can be made concerning the latter observables. 
It is well known that, at variance with the nonrelativistic calculations,
it is a common feature of the
relativistic models to overestimate the neutron skin thickness
\cite{Sug94,Sha92}. But, as seen from the results shown in Table \ref{t1},
the tensor $\rho$-N coupling has 
the tendency to correct the RMFT for this handicap.
This fact, in turn, could have very important consequences on the estimates
of the atomic parity nonconservation \cite{Pol92,Vre00}.

The GT and IA resonances in $^{208}$Pb were
computed in RRPA, for both the
NL1 and TM1 sets of parameters and within the same model space as that of
\cite{Con98}, {\it i.e.}, including only $0 \hbar \Omega$ and
$2\hbar \Omega$ excitations, and only those single-particle states
that are bound at least for neutrons.
For simplicity, we ignored the negative-energy states, although it has been 
shown \cite{Daw90} that they are required in principle even if the no-sea 
approximation is made for RMFT, since one needs a complete single-particle basis 
to develop a perfectly consistent RRPA. In fact, the transitions from Fermi- to 
Dirac-sea states are essential to ensure certain desirable features, such as 
current conservation and the removal of the spurious $J^\pi = 1^-$ translational 
state. However, such issues are not crucial for our present purposes and, 
furthermore, Ma {\it et al.} have shown in a recent calculation \cite{Ma99}
that the contribution of the negative-energy states is of decisive importance 
only for the isoscalar modes. We therefore feel safe to leave their inclusion 
for a future, more sophisticated and detailed treatment of those isovector 
resonances.

\begin{figure}[p]
\centerline{\psfig{figure=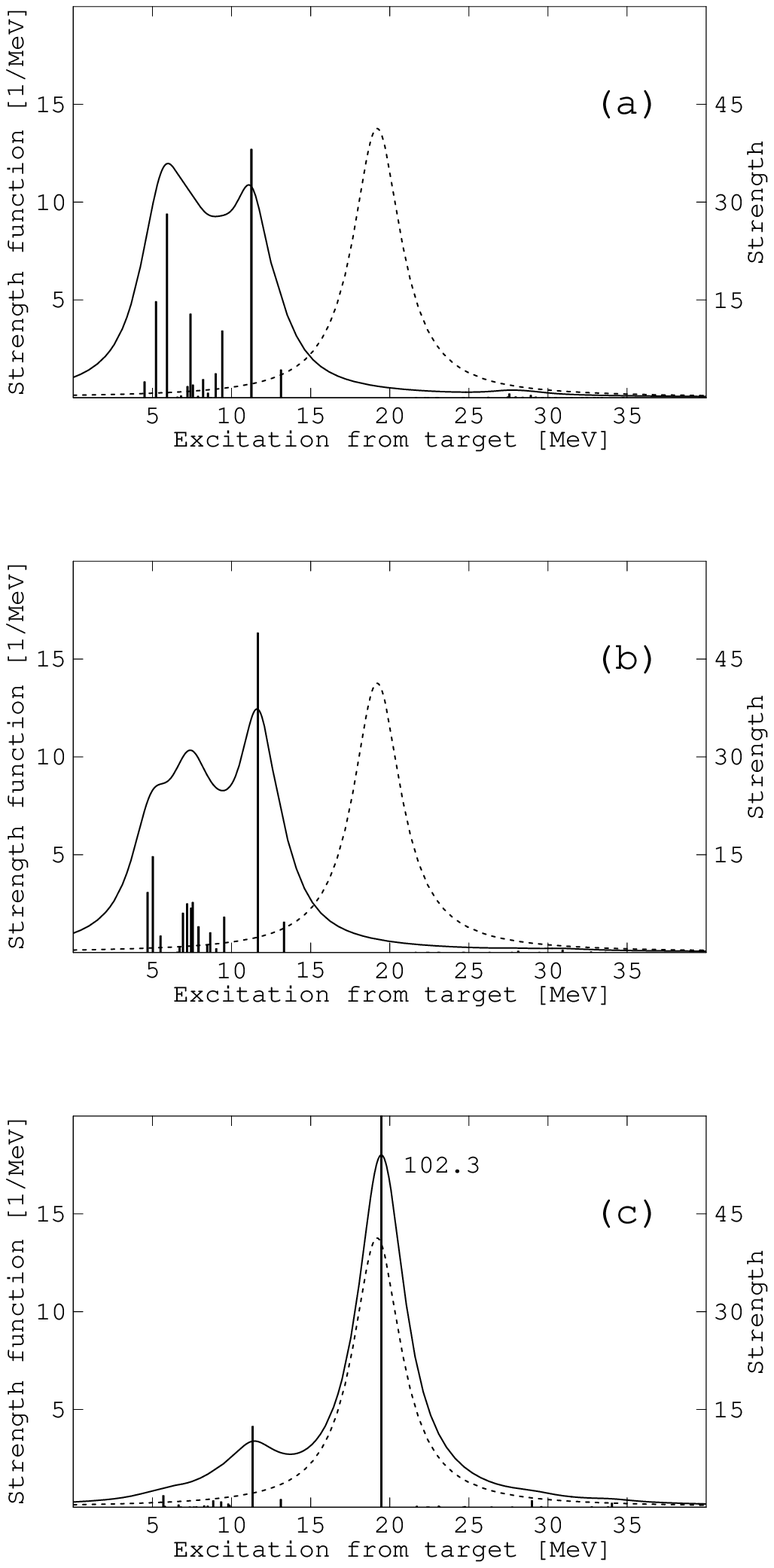,height=14cm,width=11cm,clip=}}
\caption{Gamow-Teller strength distribution for the parent nucleus
$^{208}$Pb for the parametrization NL1.
The upper, middle and lower panels correspond, respectively, to:
(a) $K_\rho=0, g'=0$;
(b) $K_\rho=3.7, g'=0$ and
(c) $K_\rho=3.7, g'=0.7$.
The spikes (r.h.s. scale) give the raw RRPA results and the continuous curve 
(l.h.s. scale), the strength function smoothed out by means of Lorentzians 
having widths of: (a) and (b) 3.0, and (c) 3.65 MeV. The strength function for 
the resonance peak extracted from experiment \protect\cite{Aki95} is drawn in 
dotted line.
\label{f1} }
\end{figure}

In Fig. \ref{f1} are shown the NL1 results for the GT strength distribution, 
both in terms of the individual strengths,  
\be
s_\lambda=\left| 
\sum_{p\bar{n}}X^\lambda_{p\bar{n}}\Bra{p}\bold{\sigma}\Ket{\bar{n}}
+\sum_{n\bar{p}}Y^\lambda_{n\bar{p}}\Bra{\bar{p}}\bold{\sigma}\Ket{n} \right|^2,
 \ee
and of a ``strength function'' obtained by  replacing the spikes by Lorentzians
of conveniently chosen widths $\Delta$ \cite{Con98}, \ie
\be
S(E)=\frac{\Delta}{\pi}\sum_\lambda\frac{s_\lambda}
{(E-E_\lambda)^2+\Delta^2}, \ee
where $X^\lambda_{p\bar{n}}$ and $Y^\lambda_{n\bar{p}}$ are,
respectively, the forward and backward going RPA amplitudes for the 
state at excitation-energy $E_\lambda$. The
upper, middle and lower panels correspond, respectively, to: (a)
$K_\rho=0, g'=0$; (b) $K_\rho=3.7, g'=0$ and (c) $K_\rho=3.7,
g'=0.7$. From these results one is induced to conclude that the
tensor $\rho$-N  coupling has a very small effect on the GT
resonance. That is, it seems as though this coupling could merely
redistribute the GT strength in the energy region between 5 and 15
MeV, but was incapable of promoting it to the correct experimental
energy. The latter is only achieved after introducing an LM force
of the same magnitude that has been used in the previous
calculation, where the just mentioned coupling has not been
considered at all \cite{Con98}. The issue of the NN-force
generated by the  $\rho$-N coupling is, however, not so simple and
deserves further discussion, which is presented below. Before proceeding, 
let us just mention that we have not noticed  large differences
between the NL1 and TM1 results for the IA and GT resonances.
For instance, in the case (c) we get that these excitations are
localized at: $E_{IA}$(NL1) = 18.6 MeV and $E_{GT}$(NL1) = 19.5 MeV
and $E_{IA}$(TM1) = 18.7 MeV and $E_{GT}$(TM1) = 20.3 MeV, while the
experimental results are: $E_{IA}$(exp) = 18.8 MeV and
$E_{GT}$(exp) = 19.2 MeV. Thus, henceforth only the parametrization
NL1 will be used.

\begin{figure}[p]
\centerline{\psfig{figure=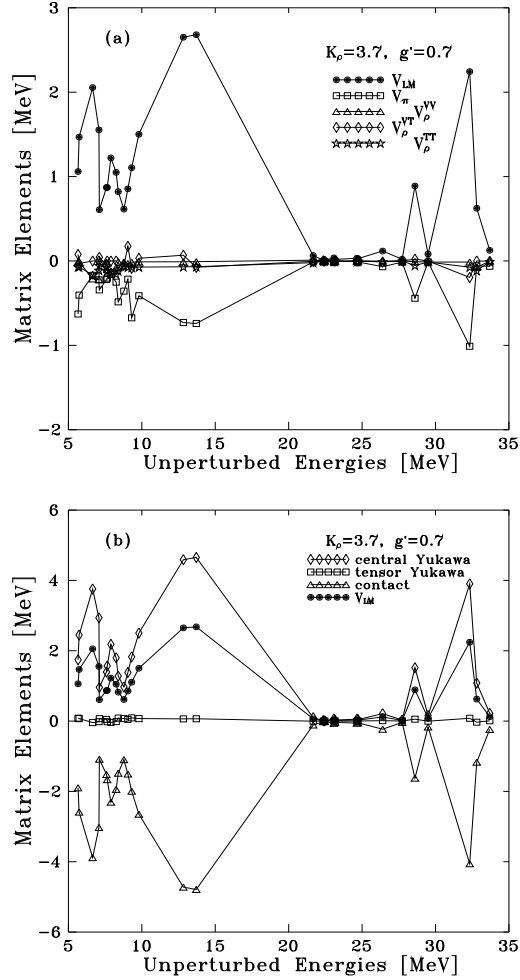,height=15cm,width=11cm,clip=}}
\caption{Diagonal matrix elements of: (a) the several terms of the
$\pi + \rho$ NN-interaction and (b) different pieces of $V_\rho^{TT}$,
taken between proton-particle neutron-hole $1^+$ states in $^{208}$Pb.
The matrix elements of the Landau-Migdal contact force are also shown in both
panels for comparison. The states are positioned at their unperturbed energies.
\label{f2} }
\end{figure}

In the upper panel of
Fig. \ref{f2} are confronted several diagonal  $J^\pi = 1^+$
proton-particle  neutron-hole matrix elements for the
$V_{LM}$, $V_\pi$, $V_\rho^{VV}$, $V_\rho^{VT}$ and $V_\rho^{TT}$ potentials.
(The meaning of the upper indices is self-explanatory.)
One can see, in particular, that the matrix elements of $V^{TT}_\rho$
are very small in comparison with those coming
from $V_{LM}$. However, when we rewrite $V_\rho^{TT}$ in the form
\begin{eqnarray}
V_{\rho}^{TT}(1,2) &=& \left( \frac{f_{\rho}}{2M} \right)^2
\bold{\tau}_1 \cdot \bold{\tau}_2 \beta_1 \beta_2 \left\{
-(\bold{\alpha}\cdot\nabla)_1 (\bold{\alpha}\cdot\nabla)_2
Y(m_\rho,r_{12})\right. \nonumber\\ &-& \frac{1}{3} m_\rho^2
\left( \frac{3}{m_{\rho}^{2}\, r_{12}^2} + \frac{3}{m_{\rho} r_{12}} + 1
\right) Y(m_\rho,r_{12}) S_{12} \nonumber\\  &+& \left.
\frac{2}{3} \left[ m_\rho^2 Y(m_\rho,r_{12}) -\delta (\bold{r}_1 -
\bold{r}_2) \right]\bold{\sigma}_1 \cdot\bold{\sigma}_{2} \right\}
 \label{10}\end{eqnarray}
and evaluate different parts separately, we find out that
the Yukawa and contact pieces in the last term engender, each one,
very large matrix elements. In fact, as shown in the lower panel, their
individual values are larger than those of $V_{LM}$, but the overall
contribution to $V_\rho^{TT}$ is small, because they cancel
each other very strongly. (A similar cancelation, though not so pronounced,
also occurs in the case of $V_\pi$.)

It should be remembered that the contact terms in  $V_\pi$ and  $V_\rho^{TT}$
would be smeared over a finite region if finite-nucleon-size effects (FNSE)
were
introduced, and they would be totally killed by realistic
short range correlations (SRC).
\footnote{Note, however, that the contributions of the contact terms are
nonzero when, both the FNSE, and the SRC
are considered simultaneously \cite{Bar99}.}
Yet, none of these two effects is considered in a mean field treatment,
such as the present one.
In return, it is common
practice \cite{Blu88,Bou87} to extract the  contact parts from \rf{8} and
\rf{9} by adding to the residual interaction the correction
term $\delta V = \delta V_\pi + \delta V_\rho\,$, with
\begin{eqnarray}
\delta V_{\pi}(1,2) &=& \frac{1}{3} \left( \frac{f_{\pi}}{m_{\pi}}
\right)^2 \bold{\tau}_1 \cdot \bold{\tau}_2 \; \bold{\sigma}_1
\cdot \bold{\sigma}_2 \; \delta (\bold{r}_1 - \bold{r}_2),
\nonumber \\ \delta V_{\rho}(1,2) &=&  \frac{1}{3} \left(
\frac{f_{\rho}}{2M} \right)^2 \bold{\tau}_1 \cdot \bold{\tau}_2 \;
\beta_1 \beta_2 \; \left( \bold{\alpha}_1 \cdot \bold{\alpha}_2 +
2 \bold{\sigma}_1 \cdot \bold{\sigma}_2 \right)
\delta (\bold{r}_1 - \bold{r}_2).
\nonumber\\
\label{11}\end{eqnarray}
For consistency, one must also perform such an extraction in the
mean field part. Since, differently from the Hamiltonian formalism
followed in \cite{Bou87}, we are working within a Lagrangian
formalism, we did this extraction in the baryon self-energy
computed in the Hartree approximation (which is equivalent to
RMFT). As a consequence the replacement $V_t\rightarrow V_t+\delta V_t$
has to be done in  the Dirac equation
\rf{6}, with
\begin{equation}
\delta V_t = \frac{1}{3} \left( \frac{f_\rho}{2M} \right)^2
\frac{\bold{\rho}_{t3} \cdot \bold{r}}{r} \tau_3
\label{12}\end{equation}
being  a correction that arises upon the extraction from the baryon
self-energy of the contact part due to this derivative coupling in eq. \rf{2}.
But, when this recipe is implemented in the numerical calculation we get
too much repulsion and the GT resonance is pushed up very high in energy.
This comes from the fact that $\delta V$ is basically a $\delta$-force of the 
type \rf{1}, with
\begin{equation}
g'_{\pi+\rho}\cong  \frac{1}{3}+
\frac{2}{3}\left(\frac{f_\rho}{f_\pi}\right)^2
\left(\frac{m_\pi}{2M}\right)^2=1.6,
\label{13}\end{equation}
which is significantly larger than $g'=0.7$.

Note that in the nonrelativistic approximation the contact term also appears
in $V_\rho^{VV}$ and $V_\rho^{VT}$, and instead of \rf{13} one would have
\begin{equation}
g'_{\pi+\rho}\cong  \frac{1}{3}+
\frac{2}{3}\left(\frac{g_\rho+f_\rho}{f_\pi}\right)^2
\left(\frac{m_\pi}{2M}\right)^2=2.3.
\label{14}\end{equation}
There is no consensus on whether one should proceed in the same way in the
relativistic case. Some authors exclude the contact terms only from
$V_\pi$ and $V_\rho^{TT}$ \cite{Bou87}, while others do that for the full
$\pi+\rho$ interaction \cite{Blu88,Bro80}. That the potentials $V_\rho^{VV}$
and $V_\rho^{VT}$
also contain a contact term follows from the substitution \cite{Bro95}
\begin{equation}
\gamma_\mu \leftrightarrow \frac{1}{2M}(2P_\mu
+\sigma_{\mu\nu}\partial^\nu)
\label{15}\end{equation}
for the vector $\rho$-N coupling.

It is worth noting that Toki and Weise \cite{Tok79} have
interpreted microscopically the LM force as arising from the
$\pi+\rho$  meson-exchange model combined with the SRC and FNSE.
In the static limit, which is used here, the result is
\cite{Ose82}:

\br
g'_{LM}(\omega=q=0)&\cong&  \frac{1}{3}
\left(\frac{  \Lambda^2 - m_\pi^2}{ \Lambda^2 + m_0^2} \right)^2
\frac{ m_0^2}{m_0^2 + m_\pi^2}
\nonumber\\
&+&\frac{2}{3}\left(\frac{g_\rho+f_\rho}{f_\pi}\right)^2
\left(\frac{m_\pi}{2M}\right)^2\frac {m^2_0}{ m^2_0+m^2_\rho },
\er
where $\Lambda$ is the cut-off mass for the pion-nucleon vertex and
$m_0^{-1}$ is the correlation length. For $\Lambda=1$~GeV, $m_0=m_\rho$
and $g_\rho+f_\rho=17.2$ this leads to $g'_{LM}(\omega=q=0)=0.67$ \cite{Ose82}.
(In the present work $g_\rho+f_\rho=23.4$.)

\vspace{\baselineskip}

Our results  can be summarized as follows:
\begin{enumerate}
\item
When the short range correlations are not considered,
the tensor $\rho$-nucleon coupling plays only a minor role
in the description of the GT resonances.
\item
If one tries to take these correlations into account by merely
extracting the contact terms from the NN interaction, the GT resonance
is pushed up too high in energy.
\end{enumerate}
Thus, the simulation of the short range correlations by the
simple-minded extraction of the contact terms alone is not a
satisfactory procedure; at least not in the case of the heavier
mesons. The explanation is that the contact terms in the $\pi +
\rho$ NN-interaction are not the only ones to be strongly modified
by the short range correlations. In particular, because of the
large $\rho$-meson mass, also the Yukawa terms generated in \rf{9}
should be strongly reduced. We conclude hence that the
implementation of, both realistic short range correlations, and
finite-nucleon-size effects, in the context of the relativistic
RPA, is required. Presently, we are working on this issue.

Finally, let us mention that the tensor $\rho$-nucleon coupling
plays an important role in the transverse spin response, and that
some progress in assessing this through a relativistic many-body 
calculation has been made quite recently  by
Yoshida and Toki \cite{Yos99}.

\vspace{\baselineskip}

The authors wish to thank Peter Ring for the use of his spherical
RMFT code, and one of us (F.K.) also thanks him for discussions
and warm hospitality at the Technischen  Universit\"{a}t
M\"{u}nchen. The work of C.D.C. was supported by CAPES
(Brazil) and FAPESP (S. Paulo, Brazil), and that of (F.K.) by
CONICET and FONCyT (Argentina) under project N$^\circ$ 03-04296.
A.P.G. and F.K. acknowledge partial financial support by ICTP
(Trieste).

%

%
%
%
%
%
%
\begin{table}[p] 
\centering
\caption{RMFT results for the energy per particle and the
root-mean-square radii of the neutron and proton point-particle
distributions in $^{208}$Pb computed with several values of the
$\rho$-nucleon tensor coupling constant $f_\rho$. The remaining
parameters are kept fixed at their NL1 and TM1 values.
\label{t1}}
\vspace{\baselineskip}
\begin{minipage}[t]{9cm}
   \begin{tabular}{lcccc} \hline \hline
   & & & &  \\
   $K_\rho\equiv f_\rho/g_\rho$ 
                   & & $E/A - M$ & $\sqrt{\langle r_n^2 \rangle}$ &
                                   $\sqrt{\langle r_p^2 \rangle}$  \\
                   & &   [MeV]   &  [fm]     &   [fm]   \\
   \hline
    & & & &  \\
     NL1 parameter set:\\
   0               & & -7.884    & 5.795     & 5.474    \\
   3.7             & & -7.882    & 5.777     & 5.480    \\
   6.6             & & -7.883    & 5.763     & 5.485   \\  & & & &  \\
     TM1 parameter set:\\
   0               & & -7.874    & 5.755     & 5.485    \\
   3.7             & & -7.871    & 5.741     & 5.492    \\
   6.6             & & -7.871    & 5.730     & 5.497   \\  & & & &  \\
   Experiment      & & -7.868\footnote{Taken from Ref. \cite{Aud97}.}
& 5.593\footnote{Taken from Ref. \cite{Bat89}.}
&5.452\footnote{Taken from Refs. \cite{Vri87} and \cite[eq. (6.1)]{Bar77}.}
   \\ \hline \hline
   \end{tabular}
\end{minipage}
\end{table}


\newpage

\begin{thebibliography}{99}

\bibitem{Ser86}B. D. Serot and J. D. Walecka, Adv. Nucl. Phys. {\bf 16} (1986)
 1.
\bibitem{Ser97} B. D. Serot and J. D. Walecka, Int. J. Mod. Phys. {\bf E 6}
(1997) 515.
\bibitem{Gam90} Y. K. Gambhir, P. Ring and A. Thimet, Ann. Phys. (N.Y.)  {\bf
198} (1990) 132, and references therein.
\bibitem{Rin96} P. Ring, Prog. Part. Nucl. Phys. {\bf 37} (1996) 193.
%
\bibitem{Blu88} P. G. Blunden and P. McCorquodale, Phys. Rev. {\bf C 38}
(1988) 1861.
\bibitem{Hui89} M. L'Huillier and N. Van Giai, Phys. Rev. {\bf C 39}
(1989) 2022.
\bibitem{She89} J. R. Shepard, E. Rost and J. A. McNeil, Phys. Rev.
{\bf C 40} (1989) 2320.
\bibitem{Daw90} J. F. Dawson and R. J. Furnstahl, Phys. Rev. {\bf C 42}
(1990) 2009.
\bibitem{Ma97} Z. Ma, N. Van Giai, H. Toki and M. L'Huillier,
Phys. Rev. {\bf C 55} (1997) 2385;
Z. Ma, H. Toki and N. Van Giai, Nucl. Phys. {\bf A 627} (1997) 1.
%
\bibitem{Con98} C. De Conti, A. P. Gale\~{a}o and F. Krmpoti\'{c}, Phys. Lett.
{\bf B 444} (1998) 14.
\bibitem{Kre81} S. Krewald, F. Osterfeld, J. Speth and G. E. Brown, Phys.
Rev. Lett. {\bf 46} (1981) 103.
%
\bibitem{Fur95} R. J. Furnstahl, H.-B. Tang and B. D. Serot, Phys. Rev. {\bf C
52} (1995) 1368; R. J. Furnstahl, B. D. Serot and H.-B. Tang, Nucl. Phys. {\bf
A 598} (1996) 539; {\bf A 618} (1997) 446.
%
\bibitem{Mac89} R. Machleidt, Adv. Nucl. Phys. {\bf 19} (1989) 189.
%
\bibitem{Con99} C. De Conti, Ph.D. Thesis, Instituto de F\'{\i}sica
Te\'{o}rica, UNESP, S\~{a}o Paulo, Brazil (1999).
%
\bibitem{Rei86} P.G. Reinhard, M. Rufa, J. Maruhn, W. Greiner and J.
Friedrich, Z. Phys. {\bf A 323} (1986) 13.
%
\bibitem{Sug94} Y. Sugahara and H. Toki, Nucl Phys. {\bf A 579} (1994) 557.
\bibitem{Bou87} A. Bouyssy, J.F. Mathiot, N. Van Giai and S. Marcos, Phys.
Rev.  {\bf C 36} (1987) 380.
%
\bibitem{Aud97} G. Audi, O. Bersillon, J. Blachot and A.H. Wapstra,
Nucl. Phys. {\bf A 624} (1997) 1.
\bibitem{Bat89} C.J. Batty, E. Friedman, H.J. Gils and H. Rebel,
Adv. Nucl. Phys. {\bf 19} (1989) 1.
\bibitem{Vri87} H. de Vries, C.W. de Jager and C. de Vries,
Atomic Data and Nuclear Data Tables {\bf 36} (1987) 495.
\bibitem{Bar77} R. C. Barrett and D. F. Jackson, {\it Nuclear Sizes and
Structure}, Oxford University Press, 1977.
\bibitem{Aki95} H. Akimune {\it et al.}, Phys. Rev. {\bf C 52} (1995) 604.
%
\bibitem{Sha92} M.M. Sharma and P. Ring, Phys. Rev. {\bf C 45} (1992) 2514.
\bibitem{Pol92} S.J. Pollock, E.N. Fortson and L. Wilets, Phys. Rev.
{\bf C 46} (1992) 2587.
\bibitem{Vre00} D. Vretenar, G.A. Lalazissis and P. Ring, nucl-th/0004018.
\bibitem{Ma99} Z. Ma, N. Van Giai, A. Wandelt, D. Vretenar and P. Ring, 
nucl-th/9910054.
\bibitem{Bar99} C. Barbero, F. Krmpoti\'c, A. Mariano and D. Tadi\'c, Nucl.
Phys. {\bf A 650} (1999) 485.
\bibitem{Bro80} G.E. Brown, {\it Lecture Notes in Physics 119}, Ed. G.E.
Bertsch and D. Kurath, Springer-Verlag 1980.
\bibitem{Bro95} L.S. Brown, {\it Quantum Field Theory}, Cambridge University
Press, 1995.
\bibitem{Tok79} H. Toki and W. Weise, Z. Physik {\bf A292} (1979) 389;
Z. Physik {\bf A295} (1980) 187.
\bibitem{Ose82} E. Oset, H. Toki and W. Weise, Phys. Reports {\bf 83} (1982)
281.
\bibitem{Yos99} K. Yoshida and H. Toki,  Nucl. Phys. {\bf A648} (1999) 75.
\end{thebibliography}
\end{document}